\documentclass[sigconf, manuscript, nonacm]{acmart}

\AtBeginDocument{%
  }
\setcopyright{acmlicensed}
\copyrightyear{2025}
\acmYear{2025}
\acmDOI{}
\acmConference[]{}{}{}

\usepackage{subcaption}
\usepackage{graphicx}
\usepackage[
    type={CC},
    modifier={by-nc-sa},
    version={4.0},
]{doclicense}

\begin{document}



\title[\textit{Negative Shanshui}]{\textit{Negative Shanshui}: Real-time Interactive Ink Painting Synthesis}

\author{Aven-Le ZHOU}
\email{aven.le.zhou@gmail.com}
\orcid{0000-0002-8726-6797}
\affiliation{%
  \institution{The Hong Kong University of Science and Technology (Guangzhou)}
  \streetaddress{No.1 Du Xue Rd, Nansha District}
  \city{Guangzhou}
  \state{Guangdong}
  \country{P.R.China}
}


\renewcommand{\shortauthors}{Zhou}

\begin{abstract}
This paper presents \textit{Negative Shanshui}, a real-time interactive AI synthesis approach that reinterprets classical Chinese landscape ink painting, i.e., shanshui, to engage with ecological crises in the Anthropocene. \textit{Negative Shanshui} optimizes a fine-tuned Stable Diffusion model for real-time inferences and integrates it with gaze-driven inpainting, frame interpolation; it enables dynamic morphing animations in response to the viewer's gaze and presents as an interactive virtual reality (VR) experience. The paper describes the complete technical pipeline, covering the system framework, optimization strategies, gaze-based interaction, and multimodal deployment in an art festival. Further analysis of audience feedback collected during its public exhibition highlights how participants variously engaged with the work through empathy, ambivalence, and critical reflection.
\end{abstract}




\begin{teaserfigure}
\centering
\begin{subfigure}{0.21\textwidth}
  \includegraphics[width=\textwidth]{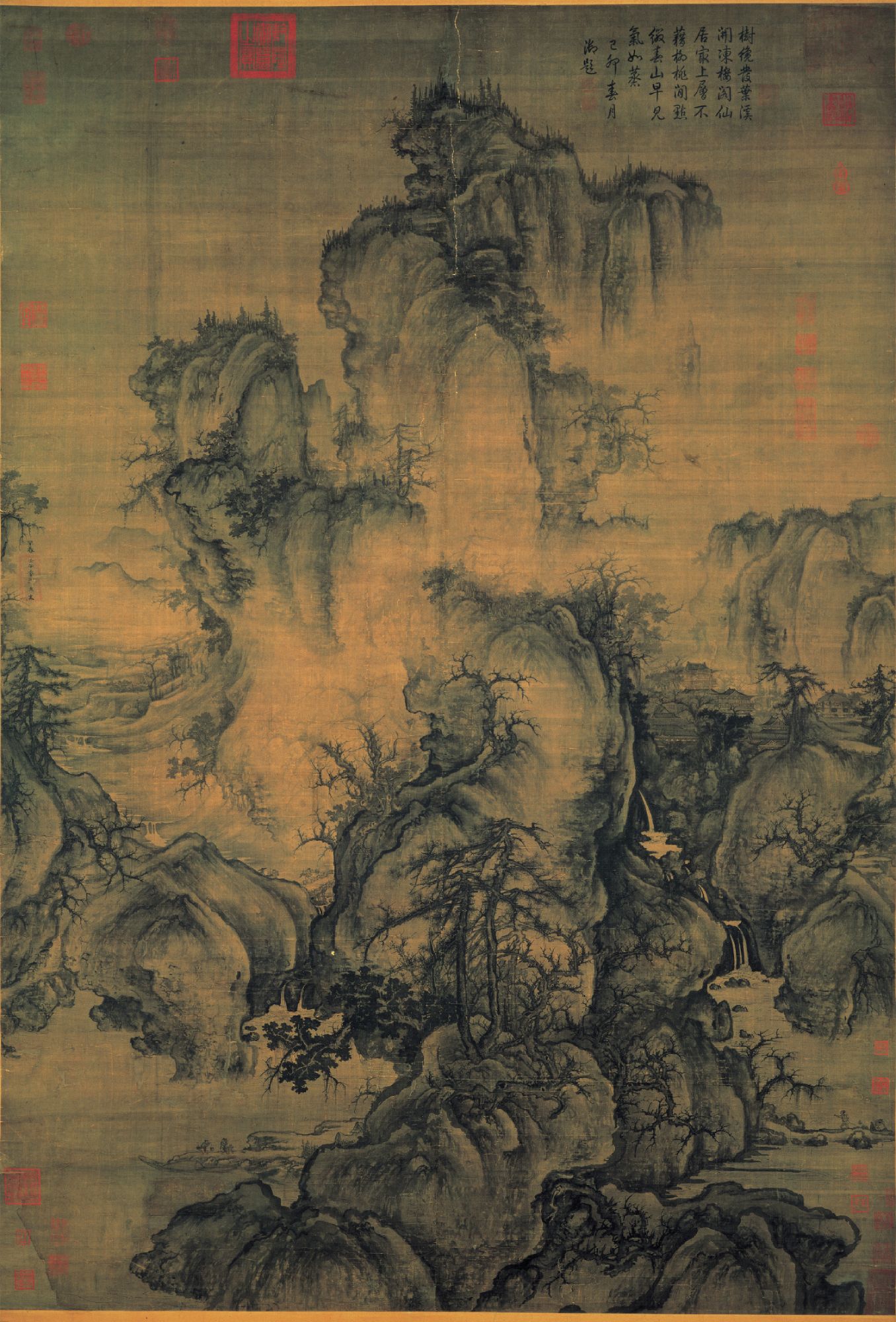}
  \caption{One Diminutive Human Walking in the Mountain.}
  \label{fig:original}
\end{subfigure}
\hfill
\begin{subfigure}{0.555\textwidth}
  \includegraphics[width=\textwidth]{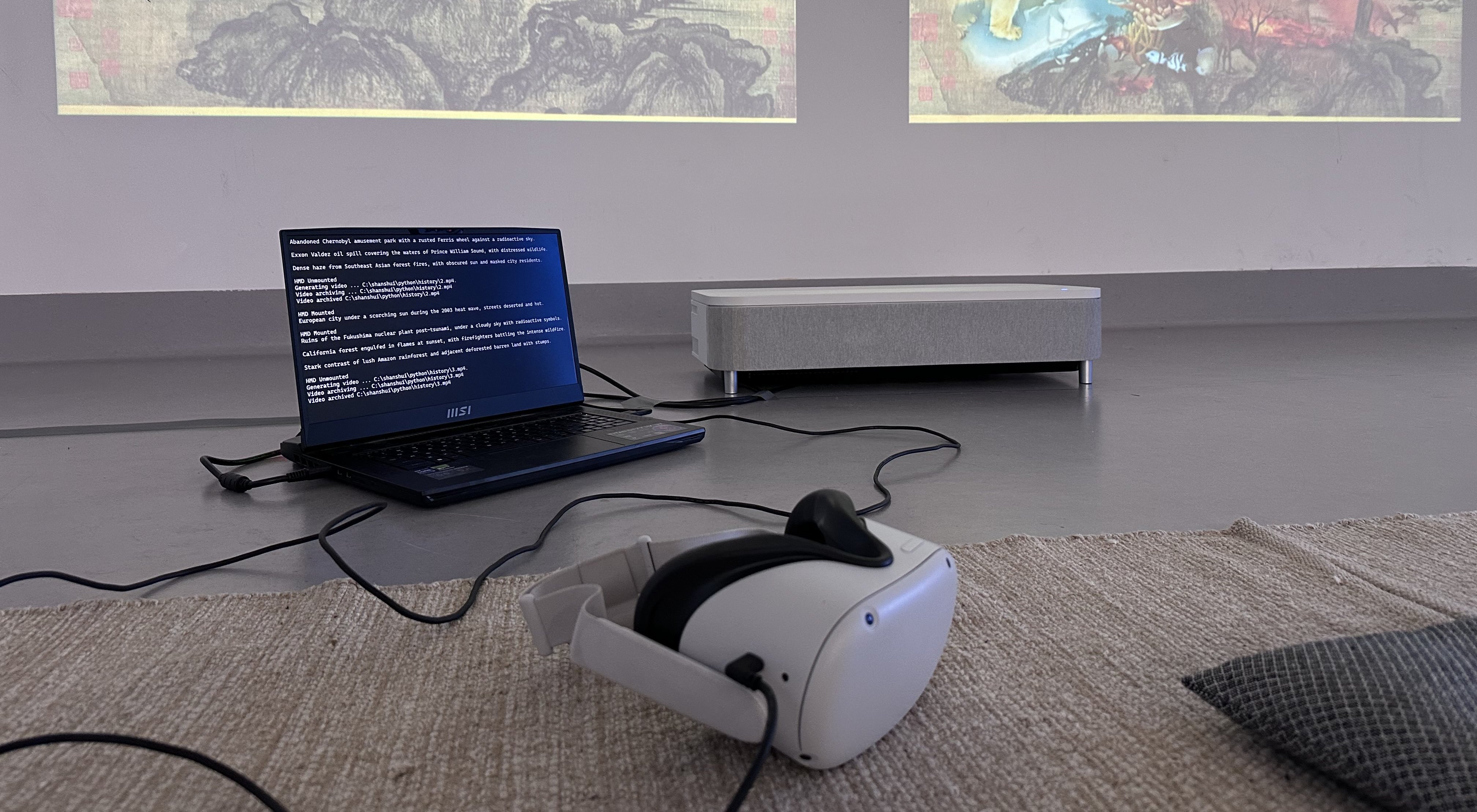}
  \caption{Integrated AI Synthesis Software and Multi-sensory Hardware System in Exhibition Setup.}
  \label{fig:equipment}
\end{subfigure}
\hfill
\begin{subfigure}{0.225\textwidth}
  \includegraphics[width=\textwidth]{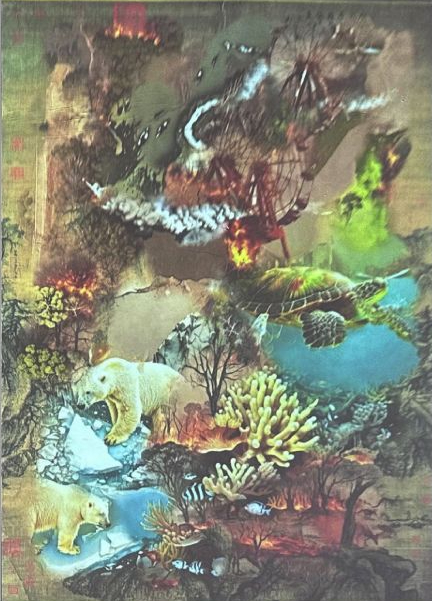}
  \caption{Synthesized Anthropocene Crisis within Shanshui.}
  \label{fig:negative}
\end{subfigure}
\caption{Real-time Interactive AI Synthesis in \textit{Negative Shanshui}.}
\label{fig:teaser}
\end{teaserfigure}

\maketitle

\section{Introduction}

In classical Chinese landscape ink painting, i.e., Shanshui, a diminutive human figure often subtly appears within the majestic mountains, reflecting an essential Eastern Daoist Oneness philosophy in the painting; this composition highlights humans as modest components within the vast natural cosmos, emphasizing humble respect for the environment. As Hui, in \textit{Art and Cosmotechnics}, profoundly points out that shanshui ``unifies the cosmic and moral orders,'' where morality arises from alignment with cosmic principles originating from the oneness \cite{hui2021art}. It represents a Daoist thought that is noninstrumental and does not seek to dominate nature. 

In contrast, contemporary society, particularly in rapidly developing economies, frequently overlooks this humility toward nature, prioritizing economic and industrial growth over environmental concerns. Human activity has now become the dominant influence on the environment, leading to interdisciplinary debates about the significant and often irreversible impact of human actions on geological and ecological systems. 

In response to the urgent environmental challenges and the conflicting interests that often delay collective action, this paper revisits the aesthetic and philosophical principles of shanshui as a lens for critical reflection. It introduces a computational approach titled \textit{Negative Shanshui}, which reinterprets traditional shanshui through real-time interactive AI synthesis. The proposed approach allows the viewer to experience a morphing transformation from serene shanshui to crisis imagery, driven by their gaze. The main contributions of this paper are:

\begin{enumerate}
    \item The design and implementation of a fine-tuned AI synthesizer integrated with a custom pipeline combining inpainting and frame interpolation to generate real-time Anthropocene crisis imagery within the shanshui painting.
    \item A complete hardware-software system enabling a gaze-responsive, multimodal virtual reality (VR) experience that supports real-time, interactive Chinese landscape synthesis and transformation.
\end{enumerate}


\subsection{Ruptured Shanshui in Contemporary Art} \label{sec:recontextrulized}

In response to the societal shift that neglects environmental respect, contemporary artists are reinterpreting shanshui to address and reflect upon the diminishing ecological concerns. Many such artworks merge shanshui aesthetics with contemporary environmental commentary; among them, artists like Yang Yongliang and Yao Lu mimic the shanshui aesthetic and contrast the serene shanshui with forged ones. 

\begin{figure}[h!]
    \centering
    \includegraphics[width=0.85\linewidth]{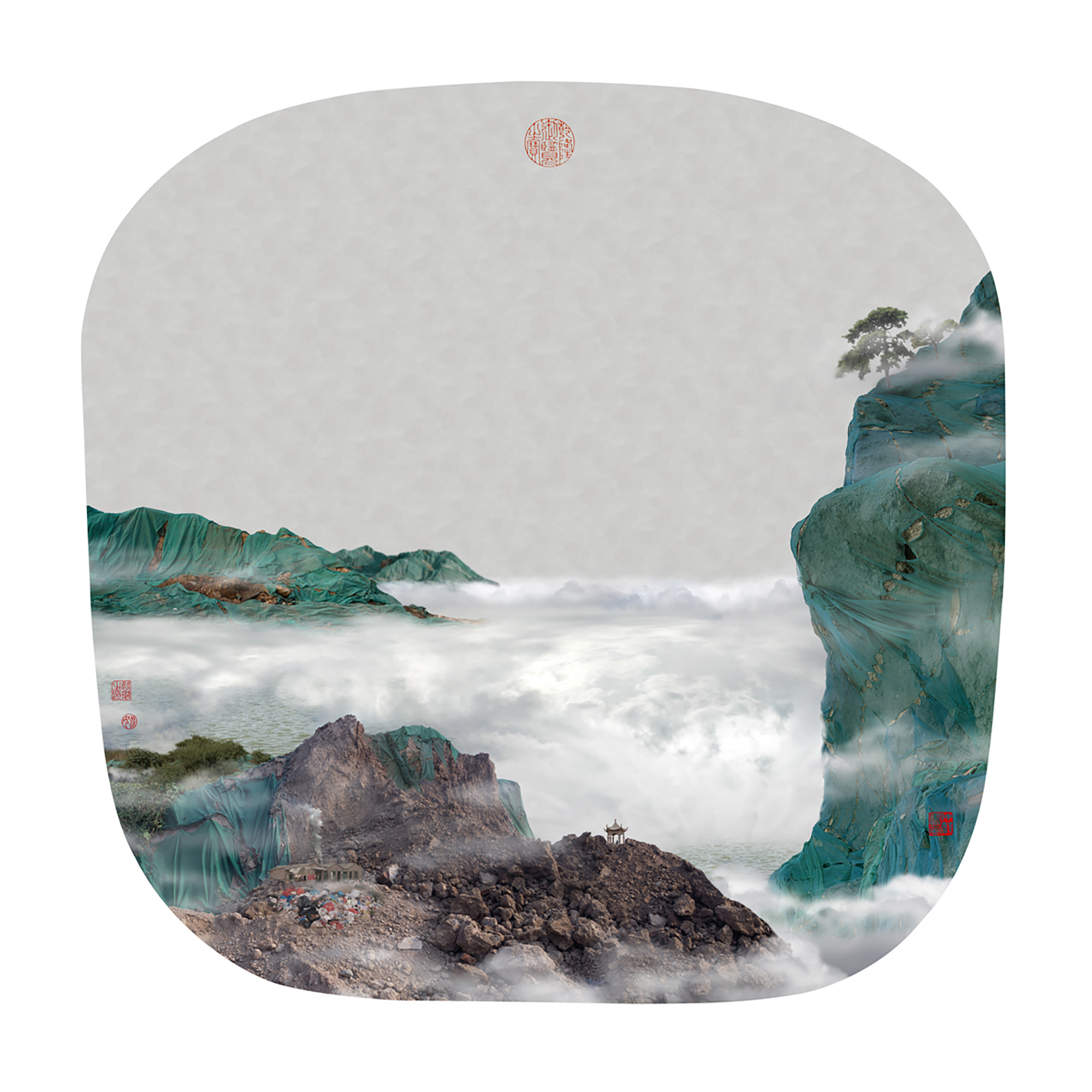}
    \caption{New Landscape Series (2009-2011) \copyright{Yao Lu}}
    \label{fig:yaolu}
\end{figure}

Yang Yongliang constructs digital landscapes where urban structures blend into natural forms, challenging viewers to discern the environmental impacts of urbanization \footnote{This can be found in his \textit{Phantom Landscape series (2006-ongoing)} at \url{https://www.yangyongliang.com/phantom-landscape/} and \textit{Artificial Wonderland series (2010-ongoing)} at \url{https://www.yangyongliang.com/new-gallery-97}}. Yang's works evolve from clear architectural influences to more subtle, natural forms, depicting the extensive impact of urbanization on nature \cite{chu_ruptured_2021}. Similarly, Yao Lu uses photographs to remix construction sites covered with green nets, commonly seen as part of China's rapid development, into serene shanshui, criticizing industrial expansion and its encroachment on natural spaces; Fig. \ref{fig:yaolu} shows his \textit{Mountain and Straw Houses in the Summer} \footnote{Part of \textit{New Landscape Series (2009-2011)} and accessible at \url{https://i.cafa.edu.cn/sub_artist/fac/show/?ai=110728&c=211&x=4&n=1371}} as an example.


These artists deliver powerful visual critiques that urge viewers to reassess the impact of urbanization and industrialization on the natural world. Upholding the aesthetic and philosophical traditions of shanshui, these artists redefine its relevance in addressing contemporary ecological challenges. \textit{Negative Shanshui} inherited this trajectory in contemporary art regarding shanshui as a framework for ecological criticism and furthered the concept with technological advancements such as AI and virtual reality (VR).

\subsection{Shanshui in the Anthropocene}
The term ``Anthropocene'' though defined in geological science, gained much more attention in social science and humanities studies \cite{adam_ditching_2024}. Nobel laureate Paul Crutzen coined Anthropocene to underscore the profound geological impact of human activities, with visible and lasting effects on Earth's ecosystems \cite{crutzen_geology_2002}. Crutzen describes this epoch as marked by the accelerated anthropogenic forces since the Industrial Revolution, which have fundamentally altered the planet's geology and ecosystems. 

Further elaborating on this epoch, Anderson critiques the current governance models that often prioritize economic and industrial growth at the expense of ecological stability, leading to widespread environmental degradation \cite{anderson_ethics_2015}. He argues for an urgent shift towards sustainability-oriented practices that effectively balance human and environmental needs. The theoretical shift aligns with the core principles of shanshui and has emerged as a focal point in contemporary shanshui discussions. 

It's evident that the conceptual transformations in the works of artists such as Yang Yongliang and Yao Lu mentioned in Sec. \ref{sec:recontextrulized} echo critiques like Anderson's about governance models that prioritize short-term economic gains over long-term ecological health. This concept behind Shanshui and Anthropocene, commonly advocating for harmonious human-nature relations, highlights the need for contemporary reinterpretations to address today's ecological challenges \cite{tan_landscape_2016} and sets the stage for our work, \textit{Negative Shanshui}. 

\subsection{\textit{Negative Shanshui} as a Critical Inquiry}

By merging the traditional ethos of Shanshui with the urgent realities of the Anthropocene, this paper critically examines the evolution of human-environment interactions. We aim to address the complexities of modern ecological issues and spark a dialogue on sustainable coexistence, emphasizing the pivotal role of Shanshui as a critical environmental inquiry in the Anthropocene. 

Our real-time interactive shanshui synthesis approach, entitled \textit{Negative Shanshui}, integrates and optimizes several components, consisting of: (1) a fine-tuned text-to-image AI synthesizer; (2) a pipeline of inpainting and frame interpretation for morphing animation generation; and (3) an integrated hardware and software system in virtual reality (VR) for real-time gaze-driven animated shanshui synthesis.

The custom AI synthesizer learns from and represents the ``negative'' ecological footprints left by humans historically in the Anthropocene. Utilizing this AI and the real-time morphing animating pipeline, \textit{Negative Shanshui} immerses the viewer in virtual reality wherein they interact with a real-time, gaze-responsive, multimodal AI system; through which the viewer transforms an aesthetic encounter with the Chinese shanshui into a critical AI-driven environmental inquiry into human–nature relations in the Anthropocene.

\section{Method}

The core narrative of \textit{Negative Shanshui} revolves around the AI-generated imagery representing the environmental degradation in the Anthropocene and its juxtaposition with original serene landscape paintings and nature. It requires three main features: (1) the theme-specific content generation and the erase, fill, and seamless merge of the generated content into the original; (2) dynamic and gradually changing animation, i.e., morphing; and (3) real-time interactive processing. 

Thus, our \textit{Negative Shanshui} approach first fine-tunes a large model with a custom text-image dataset, resulting in the AI (anthropogenic) synthesizer. It then integrates a customized pipeline of inpainting and frame interpretation to generate morphing animation. At last, it utilizes gaze interaction to drive the real-time synthesis and animated morphing to accomplish the proposed interactive experiences.

\subsection{Custom Text-to-Image Synthesis}

The AI customization process starts with collecting a comprehensive dataset of historical eco-crisis events in text and image formats, mainly through online news articles reporting significant environmental problems and events. We then fine-tune a large text-to-image model to synthesize imagery that vividly represents the theme of the custom data set. In our experiment, we use the widely adopted LoRA \cite{hu_lora_2021} technique to fine-tune Stable Diffusion \cite{Rombach_2022_CVPR} as the synthesizer of the Anthropocene crisis imaginary.

\begin{figure}[h!]
\centering
\includegraphics[width=\linewidth]{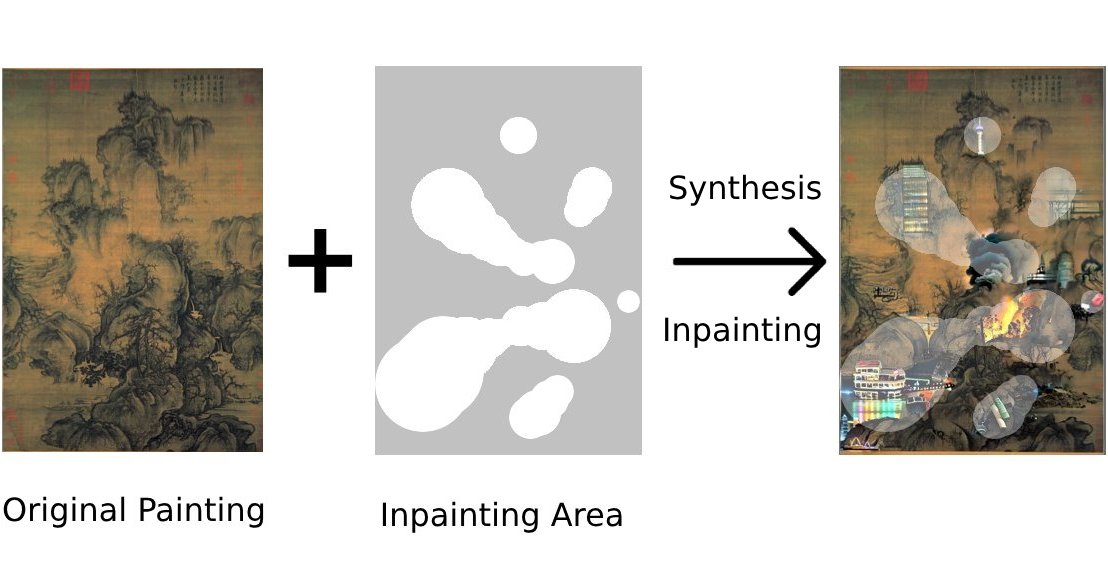}
\caption{Custom AI Synthesis \& Inpainting.}
\label{fig:area_map}
\end{figure}

\subsection{Inpainting and Frame Interpretation}

We integrate a technique known as ``inpainting'' with the AI synthesis, which allows for the erasure of specific areas designated by a user-authored mask and the filling in of these areas with AI-generated content that seamlessly merges into the original image \cite{rombach2022high}. In addition to the area mask, we use a list of ``prompts,'' to steer the thematic direction of the generated visuals. 

Building on inpainting-generated images, we apply another technique called ``frame interpretation \cite{reda2022film}'' to generate a short transitional video between each pair of adjacent frames from inpainting in real time. When sequenced, they form a continuous video stream illustrating the continuous and smooth transitions from the original image to the modified ones. 

When combining the custom AI synthesis, inpainting, and frame interpretation, it can gradually erase selected areas of the original painting, seamlessly fill with generated content, and present the process as a real-time morphing animation, making the transition from serene shanshui to the Anthropocene crisis as an ongoing narrative. 

\subsection{Gaze-based Interaction}

At last, we use the viewer's gaze to drive the real-time morphing and distortion of shanshui, of which the interaction is both natural and inevitable, enacting an ongoing engagement between the viewer and shanshui. The viewer's gaze continually drives the transformation of the shanshui world. To mimic a similar human-nature positioning in shanshui, i.e., humans as modest components within the vast natural landscape, we implement the experience in virtual reality to make a vast-scale shanshui painting possible. In the VR environment, we use ray-casting to simulate the gaze point and drive the synthesis and morphing. Combined, \textit{Negative Shanshui} situates the viewer within the real-time generated Anthropocene crisis, not as a distant observer, but as embodied within. 

\begin{figure}[h!]
    \centering
    \includegraphics[width=0.95\linewidth]{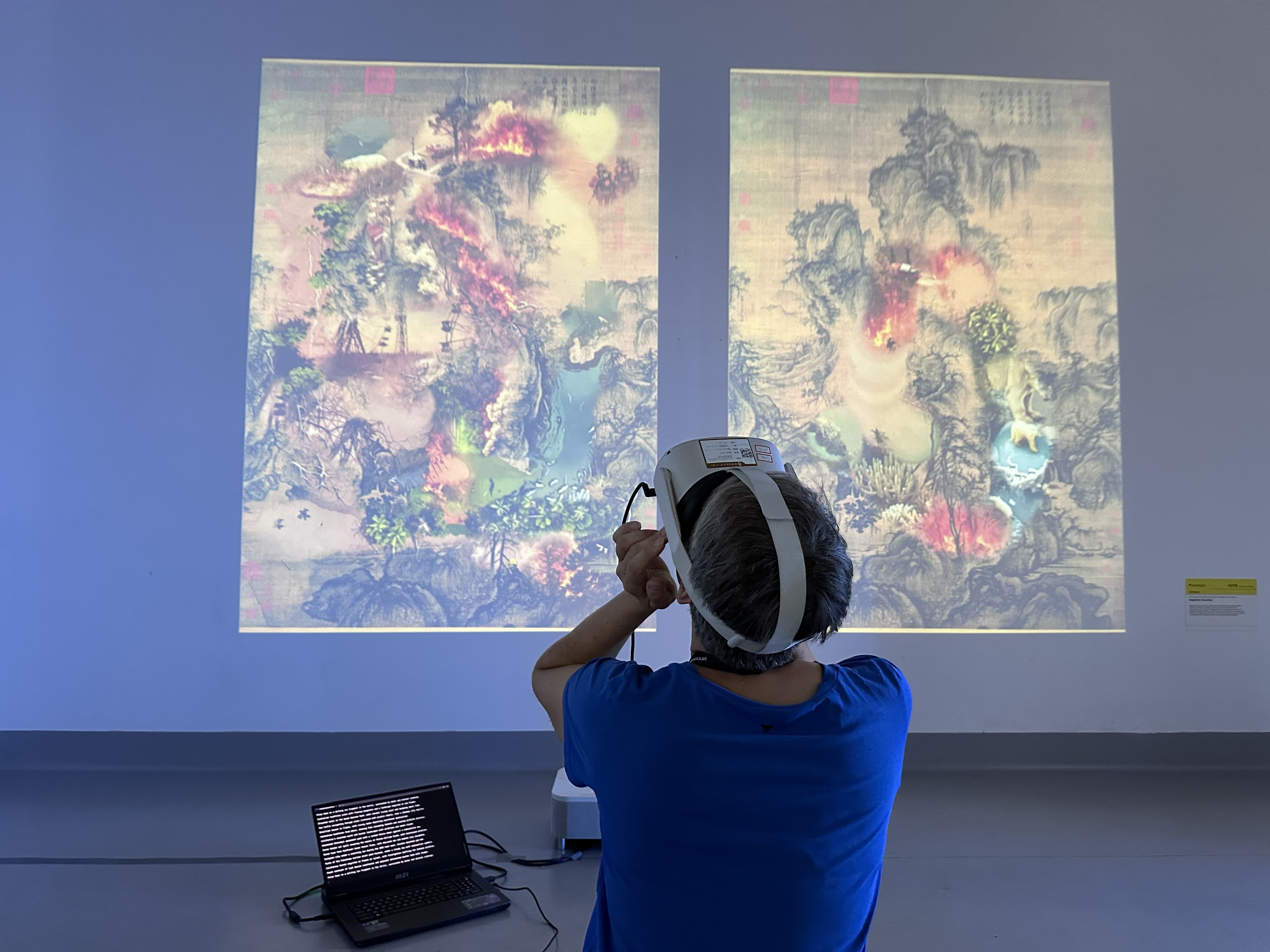}
    \caption{The Viewer Interacts with \textit{Negative Shanshui}.}
    \label{fig:enter-label}
\end{figure}

\begin{figure*}[h!]
\centering
\includegraphics[width=\textwidth]{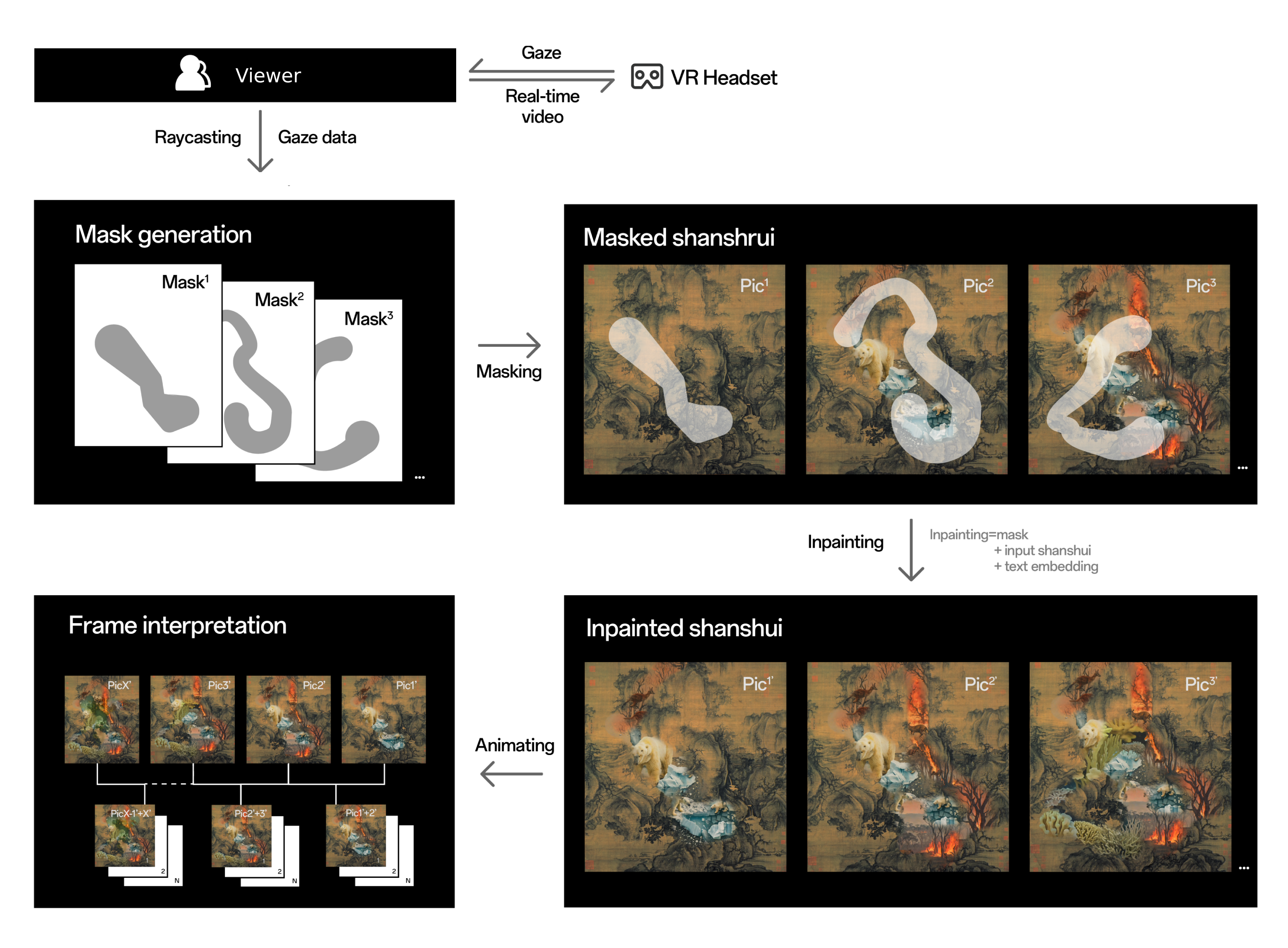}
    \caption{Technical framework in \textit{Negative Shanshui}.}
    \label{fig:architecture}
\end{figure*}

\section{The \textit{Negative Shanshui} Experiment}

In the \textit{Negative Shanshui}'s experience, the viewer wears a VR headset in which they first see a serene Shanshui masterpiece. As they view the landscape painting, their gaze is detected. The AI synthesis erases the areas the viewer gazes at and generates anthropogenic synthesized imagery, merging with the original painting. The visual progression metaphorically becomes the detrimental impacts of industrialization and consumerism on natural landscapes; The animated transitions immerse the viewer in this dramatic change and prompt them to reflect on their personal and collective roles in environmental degradation.

Through this real-time, gaze-responsive, multimodal AI system of \textit{Negative Shanshui}, the viewer transforms an aesthetic encounter with the Chinese shanshui into a critical AI-driven environmental inquiry into human–nature relations in the Anthropocene: AI-generated visuals and audio, speculative yet grounded reflections of ecological damage, merge into the shanshui world and reveal a coexistence of serenity and rupture, harmony and conflict, and landscape and degradation.

The generated imagery in VR is synchronized with a projected screen, allowing other bystanders to witness the consequences of the viewer's interaction and the resulting distortions. As part of the interactive experience, the prompt controlling the generated contents is played simultaneously as a voice-over through text-to-speech translation \footnote{Google text-to-speech API: \url{https://cloud.google.com/text-to-speech}} translation, indicating the ongoing negative actions. When the viewer removes the VR headset, the generated distortions reverse, returning to the serene Shanshui. The cyclic transformation is recorded and archived as a looping video list displayed on another screen \footnote{See a documentation video at \url{https://www.youtube.com/watch?v=KGR_AMyIlhc}}.

\subsection{System Architecture Overview} \label{subsec:architecture}

\textit{Negative Shanshui} is implemented through two synchronized software programs that support real-time interaction and generative image synthesis. The system consists of a Unity-based frontend and a Python-based backend.

The frontend is responsible for immersive rendering, gaze tracking, and bidirectional communication with the backend. It interfaces directly with the VR headset and leverages the headset's IMU (Inertial Measurement Unit) sensors to continuously capture head orientation data. From the orientation, it computes a forward-pointing vector originating at the user's virtual camera. Using ray-casting, the system determines the intersection point between this vector and the virtual shanshui canvas, which serves as the gaze coordinate. This process samples data at a frequency of approximately 70--80 points per second, providing high-resolution, low-latency tracking of the viewer's gaze across the painting. The frontend transmits these gaze coordinates in real time to the backend via a network protocol (e.g., TCP or UDP), and also monitors the headset's status---whether mounted or unmounted---sending control signals that initiate or terminate the interactive transformation process. In addition to sending data, the frontend retrieves updated image sequences from the backend and renders them to the headset display in real time, ensuring that the viewer experiences smooth, continuous morphing animations that are tightly coupled to their gaze behavior.

\begin{figure}[h!]
    \centering
    \includegraphics[width=\linewidth]{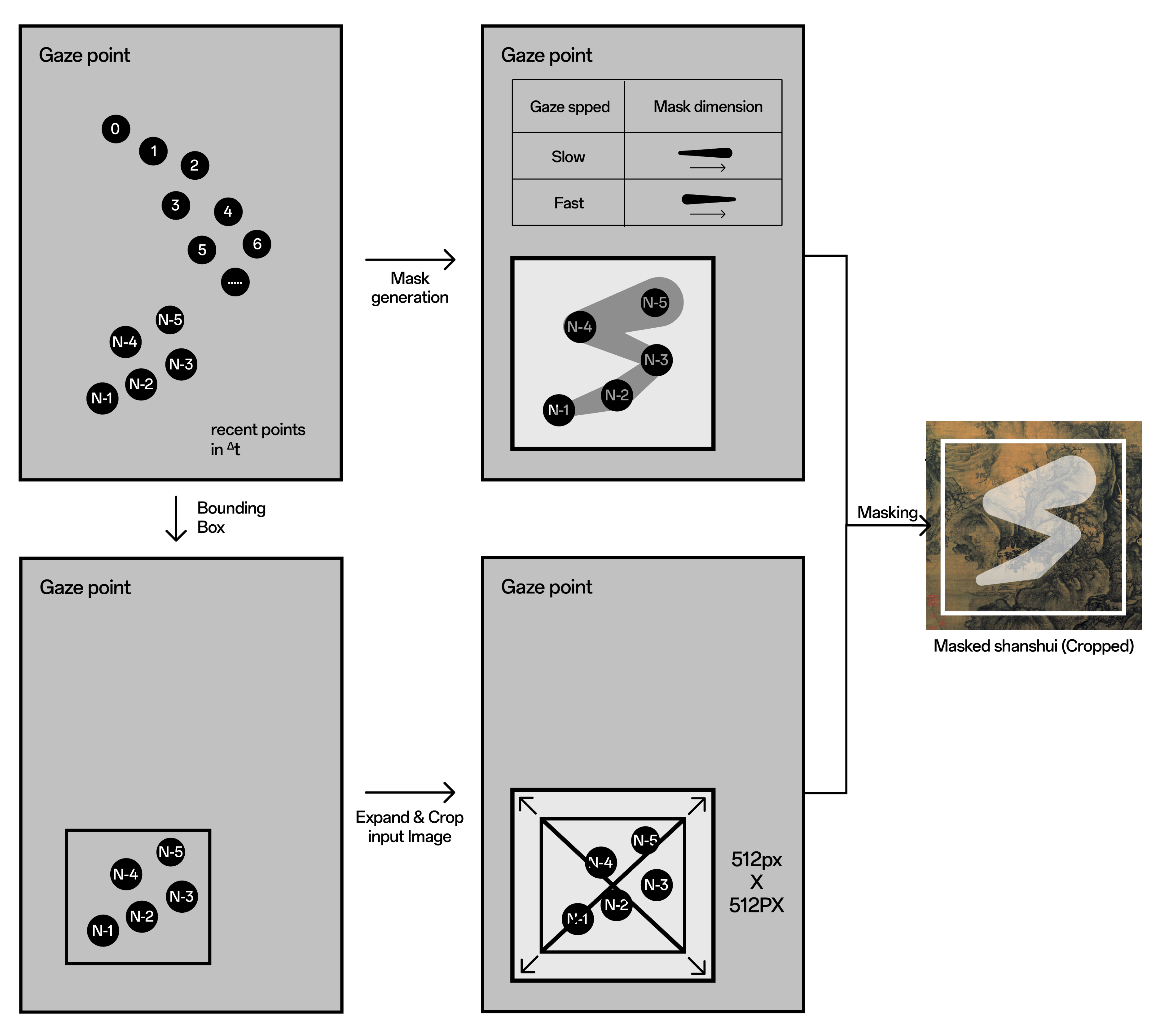}
    \caption{Mask the cropped shanshui.}
    \label{fig:spatial}
\end{figure}

The backend processes incoming gaze data through a multi-stage pipeline consisting of gaze-based masking, inpainting with Stable Diffusion, and recursive frame interpolation. The system accumulates gaze positions over a moving temporal window to generate a heatmap-style spatial mask, which defines the region of the shanshui image that will undergo transformation. The mask is computed by aggregating gaze points and connecting them with variable-width strokes that reflect gaze velocity. This mask is then used to crop a bounding box from the original high-resolution image, typically at $512 \times 512$ pixels, which is then processed for inpainting. A fine-tuned version of Stable Diffusion is employed for inpainting: the masked region is replaced with AI-generated crisis-themed imagery, guided by context-specific prompts. To optimize performance, prompt embeddings are precomputed and stored, eliminating the need to re-embed text during runtime. Inference is performed in half-precision (float16) and further accelerated using TensorFloat-32 (TF32) on compatible hardware, reducing the average inpainting time from 1.6 seconds to 0.41 seconds per frame.

Each newly inpainted image is sent to an animation module using FILM~\cite{reda2022film}, where recursive frame interpolation generates smooth transitional frames between consecutive outputs. This produces a continuous, morphing animation that responds dynamically to the user's gaze. Spatial and temporal optimizations---such as cropping only the relevant region and tuning the number of interpolated frames---ensure that the resulting animation plays at approximately 31 FPS without compromising visual quality.

The backend simultaneously streams the rendered animation sequences to two output channels: the VR headset for the immersant and an external display for bystanders. An AI-generated voiceover, produced via text-to-speech synthesis (e.g., Google TTS API), narrates contextual information that corresponds with the visual transformations, and can be routed either to the headset or to external speakers. Upon receiving a signal that the headset has been removed (``unmounted''), the backend halts the interactive session and saves the generated image and animation sequences for archival or replay.

Figure~\ref{fig:spatial} illustrates the spatial masking process, and Figure~\ref{fig:architecture} provides an overview of the full system architecture.

\subsection{Technical Challenges and Optimization} \label{subsec:challenges}

Implementing a gaze-driven, real-time AI synthesis pipeline in VR introduces several technical bottlenecks. These include image generation latency, animation coherence, gaze-to-mask stability, and system responsiveness. Each is addressed through targeted optimizations at the algorithmic and system integration levels.

\textbf{Inpainting Latency.} Generating plausible imagery using Stable Diffusion is computationally intensive. To mitigate this, we omit runtime prompt tokenization by precomputing and caching text embeddings for a curated set of thematic prompts. This eliminates redundant computation at inference time. Additionally, inference uses reduced-precision floating point (FP16) computation and TensorFloat-32 (TF32) where appropriate, accelerating performance with negligible impact on image quality. These optimizations reduce the average inpainting time from 1.6 seconds to approximately 0.41 seconds, evaluated on a high-performance GPU (RTX 3080 equivalent).

\begin{figure}[h!]
    \centering
    \includegraphics[width=\linewidth]{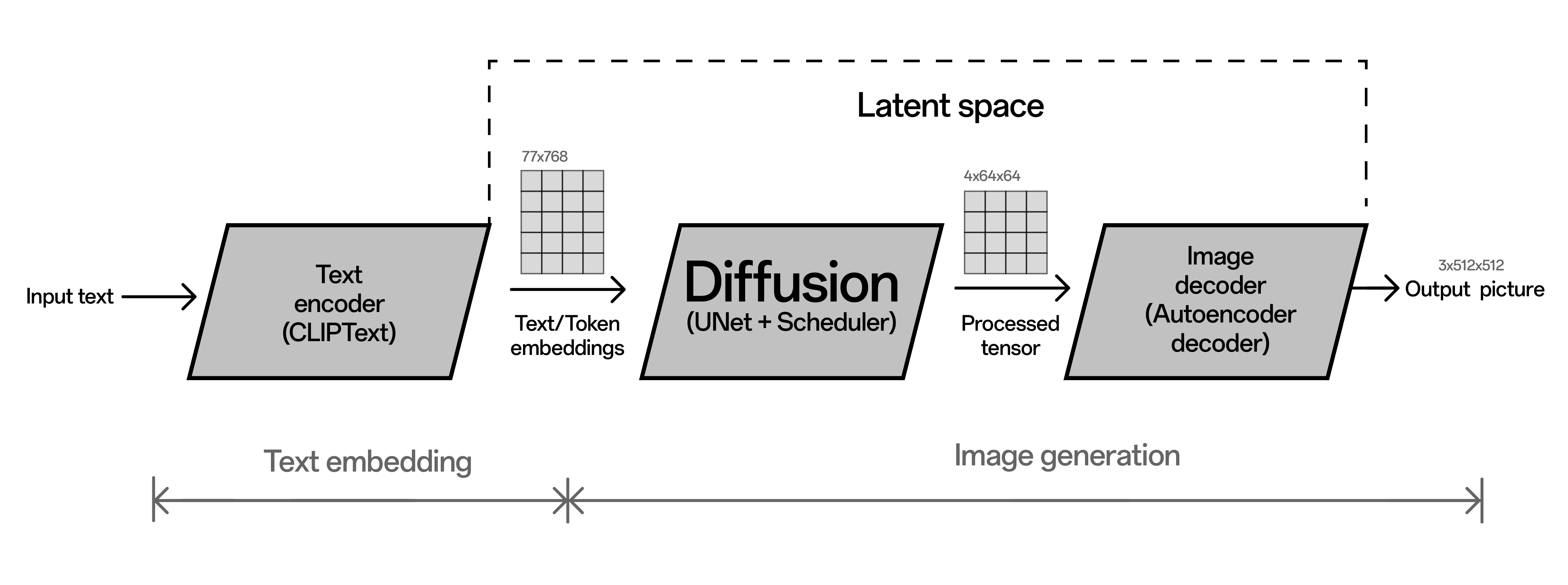}
    \caption{Omit text embedding process in runtime.}
    \label{fig:algorithmic}
\end{figure}

\textbf{Masking and Cropping Strategy.} To localize inpainting and reduce computational overhead, we implement a masking algorithm that accumulates gaze points over a temporal window $\Delta t$, typically set between 1 and 1.5 seconds. These points are connected using variable-width strokes proportional to velocity, forming a spatial mask. A bounding box is computed to enclose this region and extract a sub-image for inpainting. Given the VR canvas resolution (2250 $\times$ 1500), typical cropped regions range from 256 to 512 pixels per side. This size balances spatial granularity with processing efficiency, with fallback scaling for extreme bounding boxes.

\begin{figure}[h!]
    \centering
    \includegraphics[width=\linewidth]{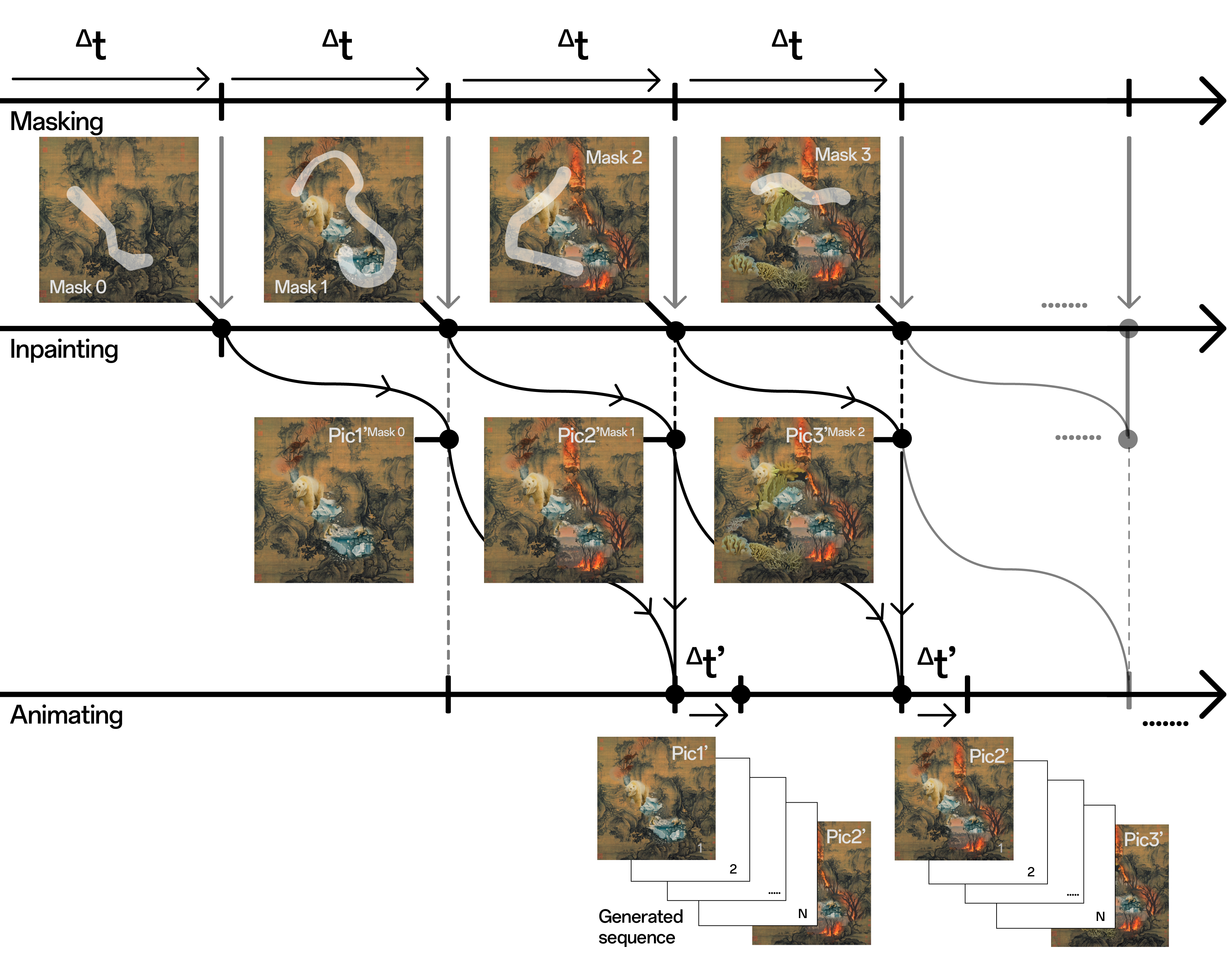}
    \caption{The optimal temporal window ($\Delta t$) for trade-off among the correlated masking, inpainting, and animating.}
    \label{fig:window}
\end{figure}

\textbf{Animation Morphing.} To ensure continuity across time, we apply a frame interpolation technique. The system generates two inpainted images spaced by $\Delta t$, then interpolates N intermediate frames using recursive interpolation. The total duration for each sequence is approximately $2\Delta t + \Delta t'$ where $\Delta t'$ is the interpolation computation time. For example, setting $N=32$ yields $\Delta t' \approx 0.21$ seconds and a smooth playback frame rate of 31 FPS. Lowering $N$ to 16 reduces computation time to 0.12 seconds but compromises smoothness.

\textbf{System Integration.} To avoid delays from synchronous communication, we implement an asynchronous messaging system between Unity and the Python backend using socket-based transmission. Gaze data and control signals are sent at ~75Hz, while image outputs are pushed to a local rendering buffer for real-time playback in VR. The backend also streams outputs to a projection screen and generates synchronized audio using a text-to-speech engine. These combined strategies enable a responsive and immersive experience where high-resolution transformations are driven fluidly by user gaze interaction.

\section{User Feedback and Analysis}

During the exhibition of \textit{Negative Shanshui} at Ars Electronica 2024, semi-structured interviews were conducted with a diverse group of visitors to explore how the work was perceived emotionally, aesthetically, and conceptually. The interview starts with two questions: ``What do you feel when you see the visuals and hear the sounds?'' and ``What's your background, and where are you from?'' Participants came from a range of backgrounds, including architecture, filmmaking, design, industrial work, and art education. While the prompt questions were intentionally open, focusing on their affective response to the visuals and sounds, and spontaneous associations, their answers revealed layered emotional and perceptual experiences. Tab. \ref{tab:audience_summary} shows our summary of the interviewee's responses, which suggest a wide spectrum of engagement, from visceral empathy to reflective distance, from playful misreadings to deeply ambivalent entanglements. In the following analysis, we identify five key thematic patterns that emerged across the interviews. These recurring modes of engagement reflect the complexity of the work's immersive design and the broader tensions of encountering mediated ecological crisis through the generative aesthetics.

\begin{table}[h!]
\centering
\renewcommand{\arraystretch}{1.35} 
\begin{tabular}{|c|p{1.6cm}|p{3.4cm}|p{1.8cm}|}
\hline
\textbf{ID} & \textbf{Background} & \textbf{Responses} & \textbf{Keywords} \\
\hline
1 & Filmmaker, South Korea & Empathy for polar bear, kangaroo, and firefighters; desire to act & empathy, urgency, emotional \\
\hline
2 & Urban Planner, Afghanistan & Neutral; compares to listening to news—continuous crisis & desensitization, media saturation \\
\hline
3 & Architect, Afghanistan & Lost in the scene, forgets nature is always present; raises question of AI metaphor & memory, nature, metaphoric AI \\
\hline
4 & Creative Coder, Spain & Feels empowered; experience fosters awareness and responsibility & agency, environmental awareness \\
\hline
5 & Artist, Cyprus & Feels like “Garden of Eden” is destroyed; visuals dominate, then sound kicks in & ruin, visual-first, sequence \\
\hline
6 & Steel Worker, Austria & Neutral; reflects duality of industry’s harm and good intentions & realism, balance, industry ethics \\
\hline
7 & Chief Design Officer, UK & Sadness, memory fades after visit; this forgetting causes more sadness & powerless, memory, forgetting \\
\hline
8 & Art Student, Germany & Initially imagines visuals from speech; finds it unexpectedly playful, draws connections to drawing & imagination, play, misalignment \\
\hline
9 & Student, Italy & Neutral news-like tone; links visuals to painting and drawing & neutrality, artistic translation \\
\hline
10 & Local, Linz, Austria & Voice reminds of Google Home; feels both domestic and distant & domesticity, mediation, distance \\
\hline
11 & (Not recorded) & No hope at first; feels guilty even looking; reversal offers emotional relief & despair, reversal, hope \\
\hline
\end{tabular}
\vspace{0.35em}    
\caption{Summary of Interviews at Ars Electronica 2024}
\label{tab:audience_summary}
\end{table}

\subsection{Empathic Immersion or Fatigue}

Several participants expressed strong emotional engagement, particularly around scenes depicting animals in crisis or human figures like firefighters. The South Korean filmmaker (P1) described feeling compassion and a desire to intervene: \textit{``If I can, I should do something.''} Similarly, the creative coder from Spain (P4) remarked: \textit{``You feel you have the power to make changes... it creates a strong feeling that we should do something.''} All suggest that \textit{Negative Shanshui} effectively encourages empathy and moral imagination, producing ethical affect through the generative aesthetics. The immersive VR experience allows viewers to witness environmental degradation and intensifies the engagement as personal and participatory.

In contrast, there are other participants who reported a sense of emotional detachment, frequently linked to the familiarity of the voiceover and soundscape. The urban planner from Afghanistan (P2) likened the experience to listening to 24-hour news: \textit{``It's like listening to news on TV, it's happening all day everywhere.''} A student in Linz (P9) emphasized the neutral tone of the standard English voiceover, while a local participant (P10) remarked: \textit{``The voice seems like my Google Home speaker... It's happening in front of me, but also distanced.''} These comments suggest that technological mediation may dull the affective immediacy of crisis imagery, reflecting what could be termed \textit{everyday crisis fatigue}, a condition where media saturation reduces emotional responsiveness.

\subsection{Temporal Disruption or Misreading}

Another group of responses pointed to a temporal dissonance: an awareness that the affective power of the experience might not endure. The designer from the UK (P7) admitted: \textit{``I feel sad... but if I go back home, I will forget about all this, which makes me feel more sad.''} Meanwhile, the Afghan architect (P3) reflected on how one could easily become absorbed in the visual transformation and lose sight of the persistent presence of nature in the background. These reflections raise questions about the work's capacity to foster \textit{durational affect}, or whether immersive experiences are ultimately ephemeral in exhibition contexts.

Some participants responded in ways that diverged from the expected emotional tone. An art student from Germany (P8) admitted to finding the experience \textit{``somewhat playful''} and was prompted to imagine drawing the visuals. Another participant (P9) similarly linked the visuals to painting. These reactions signal an interpretive ambiguity in the experience, while the subject matter leans toward environmental urgency, its aesthetic of animated painterly transformations and fluid textures invites reinterpretation. Rather than undermining the work, this playfulness could be seen as part of its speculative openness, allowing for unexpected connections.

\subsection{Dual Affective Arc: From Despair to Hope}

Perhaps the most narratively rich response came from a viewer (P11) who described a shift in emotional states across the experience. Initially overwhelmed by despair, they confessed: \textit{``I don't even know whether I should look at the beautiful parts, if I look at them, they will be destroyed.''} However, the reversal phase of the experience was described as redemptive: \textit{``The second part of reversing is much better—we need hope.''} This trajectory reflects the core dramaturgy of \textit{Negative Shanshui}, which cycles between serene shanshui aesthetics and AI-generated crisis imaginaries, and back again. The affective structure here mirrors a larger environmental psychology of grief, acceptance, and forward-looking care.


Taken together, these modes of response illuminate the multidimensional impact of \textit{Negative Shanshui}'s interactive experience and generative aesthetics for emotional, ethical, and imaginative engagements. Some participants responded with a heightened sense of agency and empathic concern, others articulated emotional saturation or distance, revealing how deeply the aesthetics of mediation and its resemblance to computational media shaped their experience. Meanwhile, spontaneous acts of imaginative reinterpretation and conflicted memory responses signal the work's openness to unpredictable meaning-making, a feature that aligns with the thematic logic of ``negative'' as both absence and generative force. 

Most notably, the affective cycle from despair to hope observed by some viewers affirms the work's structure as not only immersive but also narratively and ethically cyclical. These findings offer valuable insight into how generative, immersive art can simultaneously provoke, unsettle, and co-create meaning with its audiences, particularly when dealing with planetary-scale issues such as environmental degradation and ecological memory.

\section{Related Work}

\textit{Negative Shanshui} intersects with key developments in contemporary shanshui, generative art, immersive environmental aesthetics, and audience engagement and interaction design. Positioned at this intersection, our work draws from ecologically attuned contemporary shanshui practices while extending them through generative AI techniques to address the planetary crisis. 

This renewed engagement is evident in the works beyond the mentioned ones in Sec. \ref{sec:recontextrulized}, other contemporary artists such as Xu Bing, whose \textit{Background Stories} series (2004–ongoing) confronts the viewer with the ecological contradictions behind aesthetic conventions, transforming shanshui into a critical commentary on waste, urbanization, and environmental degradation. Other contemporary practices similarly revive shanshui as a medium for ecological reflection, situating it beyond a legacy cultural heritage but as a critical tool.

Meanwhile, a parallel wave of experimentation has emerged at the intersection of shanshui and generative artificial intelligence. Practices like Qiu Zhijie's \textit{Shanshui Spirit} (2019) and Feng Mengbo’s \textit{Wrong Code: Shanshui} (2007) explore algorithmic composition techniques rooted in classical aesthetics. However, these works often treat shanshui as a formal or stylistic template, with limited inquiry into the broader ecological or philosophical stakes of the Anthropocene. As a result, there remains a significant gap in AI-based shanshui practices that foreground environmental questions.

Rather than offering a comprehensive literature review, we highlight the following three main areas relevant to the conceptual and technical framing of \textit{Negative Shanshui}.


\subsection{Real-Time Generative Aesthetics}

From early algorithmic art to AI-based image synthesis, generative systems have enabled open-ended visual forms that foreground contingency and iteration \cite{galanter_what_2003}. Recent advancements in diffusion models and AI synthesis have introduced new possibilities for real-time generation that respond to human input \cite{zhou_steering_2025}; AI-generated ink paintings rely on various embodied interactions to engage the audience. \cite{8978182, Zhou_2019_ICCV, 10.1162/leon_a_02474} Within this lineage, \textit{Negative Shanshui} engages with generative aesthetics through a real-time interactive approach that integrates gaze tracking and image synthesis. It uses live, embodied interaction, enabling a cyclical transformation between serene shanshui imagery and animated crisis scenes. The generative aesthetics is thus situated as an engaging performative layer atop the creative interface for the viewers.

\subsection{Environmental Perception and Immersion}

Researchers have argued that sensory immersion, particularly when spatialized and dynamic, can foster deeper affective responses to planetary crisis \cite{zhou2025shanshui, demos_decolonizing_2016}. Rather than presenting environmental issues as distant spectacles, immersive media formats allow audiences to encounter layered experiences of loss, transformation, and responsibility. \textit{Negative Shanshui} builds on this trajectory by combining visual, auditory, and spatial immersion in a multimodal experience. 

The work's auditory layer, constructed in the tone of news media voiceovers, plays with the aesthetics of distance and overexposure. As audience interviews suggest, this produces a paradoxical experience: while some viewers described empathy and urgency, others interpreted the voice as desensitizing, comparable to smart home devices or broadcast media. This ambivalence aligns with existing studies on the affective complexity of climate aesthetics and media saturation \cite{yusoff_billion_2018, neimanis_bodies_2017}. In its visual composition, the work also gestures toward an aesthetic of ambiguity; the morphing imagery resists static interpretation and, at times, invites unexpected reactions, aligning with newer strands of ecological aesthetics that emphasize ambivalence, slowness, and interpretive openness \cite{sehgal_more-than-human_2024}.

\subsection{Interaction and Affective Engagement}

\textit{Negative Shanshui} also contributes to growing discourse on audience interactivity in generative and immersive environments. While participatory art often emphasizes co-creation \cite{bourriaud2002relational}, \textit{Negative Shanshui} emphasizes affective accumulation. The gaze-tracking transforms user attention into a generative force: the more the viewer looks, the more transformation and ecological disruption unfold. This model resonates with affective computing approaches that seek to embed affect into system dynamics. However, unlike many biometric or feedback-driven systems, \textit{Negative Shanshui} does not seek to understand emotions. Instead, it stages affect as a relational and cumulative process, emerging through the interplay of perception, visual transformation, and narrative.


\section{Conclusion, Limitations and Future Work}

\textit{Negative Shanshui} offers a generative, immersive experience that reimagines classical Chinese shanshui aesthetics as a critical interface for reflecting on contemporary ecological crises. Building on traditions of shanshui as contemplative encounter, the work mobilizes real-time AI synthesis and gaze-responsive interaction to provoke complex, sometimes contradictory responses from viewers, ranging from moral urgency to media fatigue, from playful reinterpretation to affective dissonance. Our analysis of audience interviews demonstrates that the proposed approach does not enforce a singular narrative but supports a plurality of affective and interpretive modes. These findings reaffirm the paper's core commitment: not to prescribe solutions to the ecological damages, but to invite engagement with the emotional and ethical dimensions of environmental perception through this traditional art form.

Meanwhile, we acknowledge the limitations of the work. While it creates a reflective and responsive environment, it remains largely centered on human perception, particularly through the use of gaze as a driver of transformation. Although effective in activating critical thought, this anthropocentric mode of interaction also underscores a broader limitation: the challenge of decentering the human while still relying on human-computer interfaces. Furthermore, while \textit{Negative Shanshui} elicits reflection, it stops short of offering pathways to action or collective ecological agency. In this sense, it foregrounds the condition of ``knowing without acting'' that Bruno Latour critiques, echoing Wang Yangming's Neo-Confucian insight that ``to know and not to act is not to know'' \cite{latour_war_2014}.

Looking forward, future iterations of the work may explore alternative interaction paradigms that reduce the centrality of human vision or incorporate non-human data inputs, e.g., environmental sensors, animal movement, or climate signals. We also see potential in expanding the aesthetic framework of shanshui beyond a medium of loss and rupture, but as a cosmotechnical model for interconnectedness, capable of expressing resilience, care, and planetary entanglement. Ultimately, the goal is not simply to use AI to visualize crisis but to foster an interactive and immersive experience to further inspire new relations between machine and human, perception and ethics, and between aesthetic encounter and ecological responsibility.
\
\bibliographystyle{ACM-Reference-Format}
\bibliography{sample-base}
\end{document}